\documentclass[final]{emulateapj}
\usepackage{apjfonts}

\usepackage{amssymb,latexsym,amsmath}

\shorttitle{Disks Around Radio Pulsars}

\shortauthors{EK\c{S}\.I \& ALPAR}

\received{2004 July 2}
\accepted{2004 Sep 13}
\begin{document}

\title{Disks Surviving the Radiation Pressure of Radio Pulsars}

\author{K. YAVUZ EK\c{S}\.{I}\altaffilmark{1}
and M. AL\.{I} ALPAR}

\affil{Sabanc\i\ University, 34956, Orhanl\i\--Tuzla, \.{I}stanbul, Turkey}

\altaffiltext{1}{Present address: Harvard-Smithsonian Center for Astrophysics, 60 Garden Street,
Cambridge, MA 02138; yeksi@cfa.harvard.edu}

\email{yavuz@sabanciuniv.edu, alpar@sabanciuniv.edu}

\begin{abstract}
The radiation pressure of a radio pulsar does not necessarily
disrupt a surrounding disk. The position of the inner radius of a
thin disk around a neutron star, determined by the balance of
stresses, can be estimated by comparing the electromagnetic energy
density generated by the  neutron star as a rotating magnetic
dipole in vacuum with the kinetic energy density of the disk.
Inside the light cylinder, the near zone electromagnetic field is
essentially the dipole magnetic field, and the inner radius is the
conventional Alfv\'en radius. Far outside the light cylinder, in
the radiation zone, $|\mathbf{E}|=|\mathbf{B}|$ and the
electromagnetic energy density is $\langle S\rangle/c \propto
1/r^2$ where $\mathbf{S}$ is the Poynting vector. Shvartsman
(1970) argued that a stable equilibrium can not be found in the
radiative zone because the electromagnetic energy density
dominates over the kinetic energy density, with the relative
strength of the electromagnetic stresses increasing with radius.
In order to check whether this is true also near the light
cylinder, we employ global electromagnetic field solutions for
rotating oblique magnetic dipoles (Deutsch 1955). Near the light
cylinder the electromagnetic energy density increases steeply
enough with decreasing $r$ to balance the kinetic energy density
at a stable equilibrium. The transition from the near zone to the
radiation zone is broad. The radiation pressure of the pulsar can
not disrupt the disk for values of the inner radius up to about
twice the light cylinder radius if the rotation axis and the
magnetic axis are orthogonal. This allowed range beyond the light
cylinder extends much further for small inclination angles. The
mass flow rate in quiescent phases of accretion driven millisecond
pulsars can occasionally drop to values low enough that the inner
radius of the disk goes beyond the light cylinder. The
possibilities considered here may be relevant for the evolution of
spun-up X-ray binaries into millisecond pulsars, for some
transients, and for the evolution of young neutron stars if there
is a fallback disk surrounding the neutron star.
\end{abstract}


\section{Introduction}

A neutron star can interact with a surrounding disk in a variety
of modes. These different modes of interaction can lead to various
astrophysical manifestations and are of key importance in
classifying neutron stars. The mode of interaction and hence the
evolutionary stage is determined by the location of the inner
radius of the disk with respect to the characteristic radii, the
corotation radius $R_c=(GM/\Omega^2)^{1/3}$ and the light cylinder
radius $R_L=c/\Omega$ where $\Omega$ is the angular velocity and
$M$ is the mass of the neutron star. Starting with the pioneering
work of \citet{shv70a}, three basic modes of interaction
--accretor, propeller and ejector-- \citep{lipunov} of a neutron
star with a surrounding disk have been identified. If the inner radius of the disk
is beyond the corotation radius, the system is expected to be in the 
propeller stage \citep{shv70a,IS75,DO73,fabian}. 
The ejector (radio pulsar) stage is assumed to commence
when inner radius is also beyond the light cylinder radius.

The disk should be disrupted in the region where magnetic
and matter stresses are comparable \citep{PR72,DO73,LPP73}
and the
inner radius of the disk, $R_{in}$, is estimated by balancing
the kinetic energy density
(half the ram pressure) of the disk with the
electromagnetic energy density (same as electromagnetic pressure)
of the radiation generated by the neutron star. In order that the
equilibrium found in this way is stable, the electromagnetic
energy density at $r<R_{in}$ should be greater than the kinetic energy density, i.e.\
the electromagnetic energy density at $R_{in}$ should increase more steeply than the
kinetic energy density in approaching the star \citep{lipunov}. The usual
estimate for the kinetic energy density has the dependence
$\mathcal{E}_{K} \propto r^{-5/2}$ (see \S 3) on the radial distance $r$
  from the center of the
neutron star. In the near zone ($R_{\ast}<r \ll R_L$) where
$R_{\ast}$ is the radius of the neutron star, electric energy
density, $\mathcal{E}_{e}$, is negligible and the kinetic energy
density is balanced essentially by the magnetic energy density,
$\mathcal{E}_{m}$, of the dipole field of the neutron star at the
Alfv\'en radius. This is a stable equilibrium point because
$\mathcal{E}_{em} \propto r^{-6}$ is steep enough to balance
 $\mathcal{E}_{K} \propto r^{-5/2}$ i.e.
$\mathcal{E}_{em}>\mathcal{E}_{K}$ for $r<R_{in}$ and $\mathcal{E}_{K}>\mathcal{E}_{em}$
for $r>R_{in}$.
In the radiation zone ($r \gg R_L$)
where electromagnetic energy density $\mathcal{E}_{em} \propto r^{-2}$,
a stable inner boundary can not be
found because then $\mathcal{E}_{K}$ would be greater than $\mathcal{E}_{em}$
for $r<R_{in}$ and $\mathcal{E}_{em}>\mathcal{E}_{K}$ for $r>R_{in}$, and the disk
is disrupted.
This was first noticed by \citet{shv70b} who with the above
argument concluded that the stable equilibrium outside the light
cylinder would only be possible  beyond the gravitational capture
radius \citep{lipunov}.

\begin{figure}[t]
\epsscale{1.0} 
\plotone{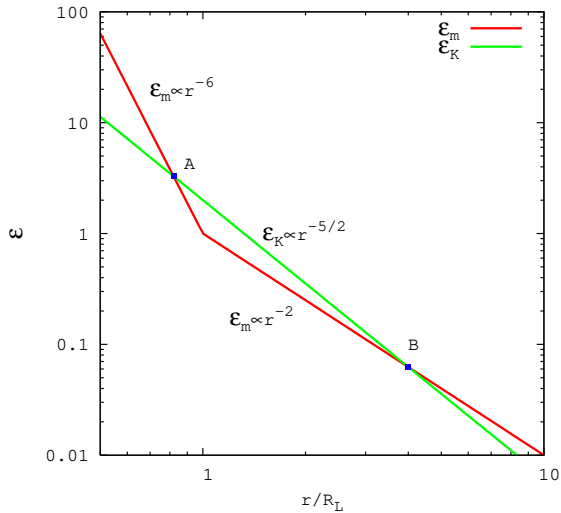}
\caption{Two possible equilibrium configurations around a rotating dipole. 
Equilibrium at the near zone (point A = Alfv\'en radius) is stable because magnetic energy density 
drops more rapidly than the kinetic energy density going beyond this point.
Equilibrium at the radiation zone (point B) is not stable because the magnetic pressure drops
less rapidly than kinetic energy density and disrupts the disk beyond this point. With a piecewise 
defined field configuration as above, one has to conclude that the disk has to be disrupted
once the inner radius goes beyond $R_L$. (\emph{This figure is not included
in the accepted paper. It is added for explanatory purposes.})}
\label{extra}
\end{figure}

Usual estimates employ a rotating magnetic dipole in vacuum to generate the
electromagnetic fields of the neutron star. In the near zone
 ($R_{\ast}<r \ll R_L$) the magnetostatic energy density $B^2/8\pi$ is
employed, while in the radiation zone ($r \gg R_L$), at distances
far outside the light cylinder, the radiation pressure $\langle S
\rangle /c$, where $\mathbf{S}$ is the Poynting vector, is used.
For simplicity one usually employs a piecewise defined
electromagnetic energy density which scales as $r^{-6}$ for
$r<R_L$ and as $r^{-2}$ for $r>R_L$. With such a model one
necessarily concludes that the disk will be disrupted if the inner
radius of the disk goes beyond the light cylinder radius (see Figure~\ref{extra}).

In this work we present the full expressions for electromagnetic
pressure around a rotating dipole in vacuum, reducing to the
conventional expressions in the near zone and the radiation zone,
and describing the transition across the light cylinder. We use
these results to investigate the location and stability of the
Alfv\'en radius and inner radius of thin Keplerian disks. This
information is relevant if the mass flow rate through the disk
declines or the neutron star rotation period evolves such that the
inner radius is beyond the light cylinder. According to the conventional 
pulsar magnetophere models \citep{GJ69}, the pulsar activity can commence
if the inner radius is outside the light cylinder. According to the 
simple piecewise model described above, the disk would be swept away. 
In the present model with a rather broad transition from the near zone 
to the radiation zone, there will be a wide range of allowed values 
for the mass flow rate for which the inner radius of the disk is 
outside the light cylinder while the disk can survive the radiation pressure.
In this case, the pulsar activity will commence 
but depending on the inclination angle, the presence of the disk
may destroy the coherent pulsed emission in the radio band.

\begin{figure}[th]
\epsscale{1.0} 
\plotone{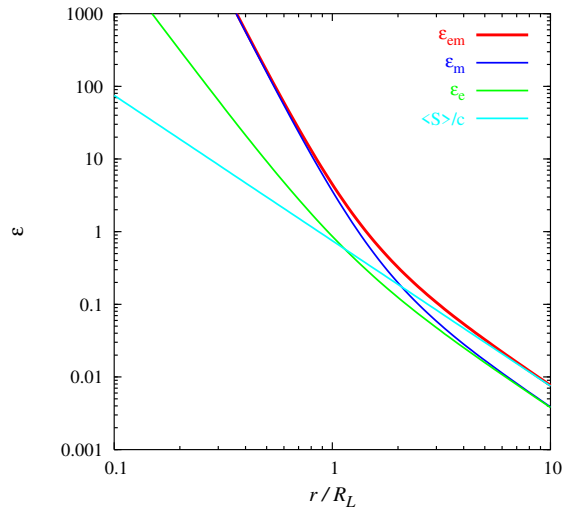} \caption{Radial structure of the
electromagnetic energy
density around the pulsar. \emph{Solid} (colored red in electronic edition) line is the
electromagnetic energy density, \emph{long dashed} (colored green in electronic edition) 
line is the electrical energy density $E^2/8\pi$, \emph{short dashed}
(colored dark blue in electronic edition) line is the magnetic energy density $B^2/8\pi$, and
\emph{dashed and dotted} (colored light blue in electronic edition) line is
$\langle S_r\rangle /c$. Magnetic energy density dominates
inside the light cylinder radius. Far outside the light cylinder
electric and magnetic energy densities become equal and
$\langle S_r \rangle /c$ is equal to their sum. 
See the electronic edition of the Journal for a color version of this figure.
\label{main}}
\end{figure}

\placefigure{main} 

In its quiescent stage,
the luminosity of the accretion driven millisecond pulsar
\objectname{SAX J1808.4-3658} \citep{wijnands98} drops below 
$L_x \sim 10^{32}$ erg s$^{-1}$ \citep{campana02}. Assuming the system passes 
through a propeller stage the mass inflow rate, which determines the 
inner radius, can be much greater than 
the mass accretion rate on to the star determining the luminosity. 
As we do not know what fraction of the inflowing mass can accrete in the propeller
stage, we can not estimate the inner radius from the observed luminosity. 
Assuming 1 per cent of the inflowing mass accretes on to the star, we would estimate 
an inner radius for the disk greater than the light cylinder. 
\citet{burderi03} suggested that the optical 
properties of this source in the quiescent stage indicate an active
pulsar (see also \citet{campana04}). We find that such a disk can
survive the radiation pressure of a 
turned on rotation-powered pulsar. This may be the case in some stages 
of transients (e.g. \citet{campana,zamanov}) and of the evolution of young neutron
stars if they have fallback disks
\citep{MD81,marsden01b,marsden02,menou01,AAY01,EA03}.

In \S 2 we derive the electromagnetic energy density from the
global electromagnetic field solution of \citet{deu55} for
obliquely rotating  magnetic dipoles. In \S 3 we derive the inner
radius of the disk. In \S 4 we
discuss implications for the accretion driven millisecond pulsar
\objectname{SAX J1808.4--3658} which is probably representative 
of the millisecond X-ray pulsars, for certain other transients and for fallback disks around
radio pulsars. In \S 5 we discuss our results.

\section{Electromagnetic Energy Density}

A rotating dipole can radiate electromagnetic energy if the spin
and magnetic axes are not aligned. A global solution for a
perfectly conducting, rigidly rotating star in vacuum was first
given by \citet{deu55}. The relevance of this solution for radio
pulsars is thoroughly discussed by \citet{ML99}.

The early papers on pulsars \citep{OG69,pacini67,pacini68,MG70,DG70} 
employed the vacuum solutions of \citet{deu55}. The vacuum assumption 
can not actually be realized for neutron stars because the magnetospheric field 
itself is strong enough to tear off electrons from the surface of the neutron 
star. Indeed, \citet{GJ69} showed that an axisymmetric pulsar magnetosphere 
should have a corotating plasma described by the hydromagnetic approximation
$\mathbf{E}+\mathbf{v\times B}=0$ with a corresponding charge density 
$\mathbf{\nabla \cdot E}$. Whether the pulsar electrodynamics
is dominated by the \citet{GJ69} currents, or by the vacuum fields
modified by rather smaller currents is still of debate \citep{mich91} and a global
self-consistent solution of the problem is yet to be presented.
\citet{kaburaki81} argued that the existence of a corotating
plasma does not alter the field structure drastically because it
simply means an effective increase in the radius of the conducting
star (see \citet{melatos97} for an application of this idea)
using the \citet{deu55} solution. The presence of the disk might alter
the field structure (e.g. \citet{EL77}) even more than the corotating plasma and further
complicate the magnetospheric currents \citep{treves}. In the absence of a 
consistent solution of the problem of the pulsar magnetosphere with corotating 
plasma with or without a disk, we, for simplicity, employ the \citet{deu55} solutions
for a rotating dipole in vacuum.
We quote the field solutions of \citet{deu55}, as corrected by \citet{ML99}, 
in the Appendix, and use them in our estimation of the electromagnetic energy density
at the disk plane. As Maxwell's equations do not preserve their form in rotating 
coordinates the problem of rotating dipoles is intrinsically relativistic. These 
solutions can be shown to be the limiting case  for slow rotation (see e.g. \citet{belinski}).

As obtained in the Appendix, the radial component of the Poynting
vector $\mathbf{S}=(c/4\pi ) \mathbf{E\times B}$ at the disk plane
(zenith angle $\theta=\pi/2$) averaged over a stellar period is
\begin{equation}
 \frac{\langle S_r \rangle}{c} =\frac{\mu ^{2}}{8\pi R_{L}^{6}}
\frac{\sin ^{2}\xi }{x^{2}} \label{SC}
\end{equation}
Here $\xi$ is the angle between the magnetic moment $\mathbf{\mu}$
and the rotation rate $\mathbf{\Omega}$ of the neutron star, and
$x=r/R_L$. Note that Eq.~(\ref{SC}) is
accurate only for $\alpha=R_{\ast}/R_L \ll 1$. For the fastest
millisecond pulsar with $P\cong 1.5$ ms, $\alpha=0.14$, thus
$\alpha \ll 1$ holds for pulsars; in this case
$\langle S_r \rangle /c  \propto r^{-2}$ for $r$ greater than a few
$R_{\ast}$, (see Eqs. (\ref{S_r_App})-(\ref{As})).

The dynamical disk time-scale is much longer than the period of
the neutron star. The disk would only ``see'' the average field of
many stellar rotations. We average the squared fields over one
stellar period $P$ as $\langle B_{i}^{2}\rangle
=(1/P)\int_{0}^{P}B_{i}^{2}dt $ where $B_{i}$ is any component of
the magnetic field, and obtain the magnetic energy
density $\mathcal{E}_{m}=\langle B^{2}\rangle/8\pi$ as (see Appendix)
\begin{equation}
\mathcal{E}_{m} \simeq \frac{\mu ^{2}}{8\pi R_{L}^{6}}x^{-6}\left[ \cos ^{2}\xi +\frac{1}{2}\left(
x^{4}+3x^{2}+5\right) \sin ^{2}\xi \right] \label{eq_Em}
\end{equation}
for $\alpha \ll 1$.

\begin{figure}[th]
\epsscale{1.0} 
\plotone{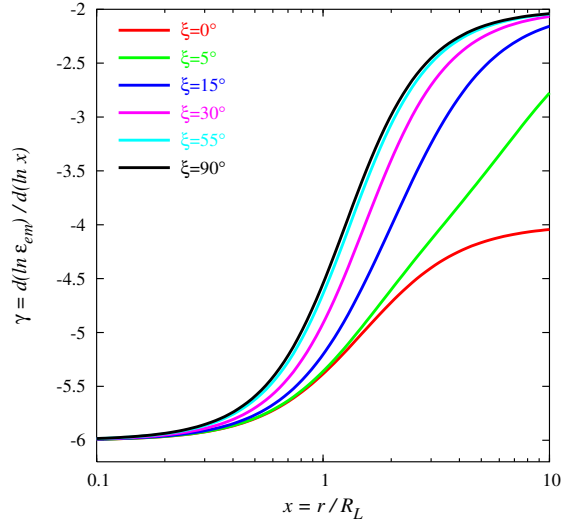} \caption{Power-law index of the
electromagnetic energy density for a variety of
inclination angles. The power-law index changes between -6 and -2.
The transition becomes broader for small inclination angles.
See the electronic edition of the Journal for a color version of this figure.
\label{gamm1_fig}}
\end{figure}

\placefigure{gamm1_fig}  

Similarly, in the $\alpha \ll 1$ approximation, the electric energy density
$\mathcal{E}_{e}=\langle E^{2}\rangle/8\pi$  is (see Appendix)
\begin{equation}
\mathcal{E}_{e}\simeq \frac{\mu ^{2}}{8\pi R_{L}^{6}} x^{-4}\left[
\frac{4}{9}\cos ^{2}\xi +\frac{1}{2} \left(x^{2}+1\right) \sin
^{2}\xi \right] \label{eq_Ee}
\end{equation}
The total electromagnetic energy density can be found by summing up
the magnetic and
electric energy densities. From equations (\ref{eq_Em}) and (\ref{eq_Ee}), one finds
\begin{equation}
\mathcal{E}_{em}=\frac{\mu ^{2}}{8\pi R_{L}^{6}}x^{-6}\left[ \left( 1+\frac{4}{9}x^{2}\right)
\cos ^{2}\xi +\left( x^{4}+2x^{2}+\frac{5}{2} \right) \sin ^{2}\xi \right]
\label{E_tot2}
\end{equation}
Note that this reduces to $\mathcal{E}_{em} \propto r^{-6}$ for
$r \ll R_L$ and $\mathcal{E}_{em} \propto r^{-2}$ for $r \gg R_L$,
as expected.
We show $\mathcal{E}_{em}$ together with $\mathcal{E}_{m}$,
$\mathcal{E}_{e}$, and $\langle S_r/c\rangle$, all scaled with
$\mu ^{2}/8\pi R_{L}^{6}$, in Figure~(\ref{main}). If the sole
effect of the corotating plasma is to increase the radius of the
region beyond which vacuum starts from $R_{\ast}$ to a larger
value $R_v$ \citep{melatos97}, then $\alpha\equiv R_v/R_L$ can have
values close to unity. In this case the expression for the total
energy density is more complex than equation~(\ref{E_tot2}), the coefficients
having terms with powers of $\alpha$. The
behavior of $\mathcal{E}_{em}$ and its constituents do not change
qualitatively for $\alpha \lesssim 1$ if we retain such terms.


The local power-law index of electromagnetic energy density,
$\gamma \equiv d\ln \mathcal{E}_{em}/d\ln x$, calculated from equation (\ref{E_tot2}) is
\begin{equation}
\gamma=-6+\frac{(8/9)x^{2}\cos
^{2}\xi +4x^{2}\left( x^{2}+1\right) \sin ^{2}\xi }{\left[ 1+(4/9)
x^{2}\right] \cos ^{2}\xi +\left( x^{4}+2x^{2}+5/2\right) \sin ^{2}\xi } \label{eq_gamm}
\end{equation}
We show $\gamma $ for a variety of $\xi $ in Figure~(\ref{gamm1_fig}).

\begin{figure}[h]
\epsscale{1.0} 
\plotone{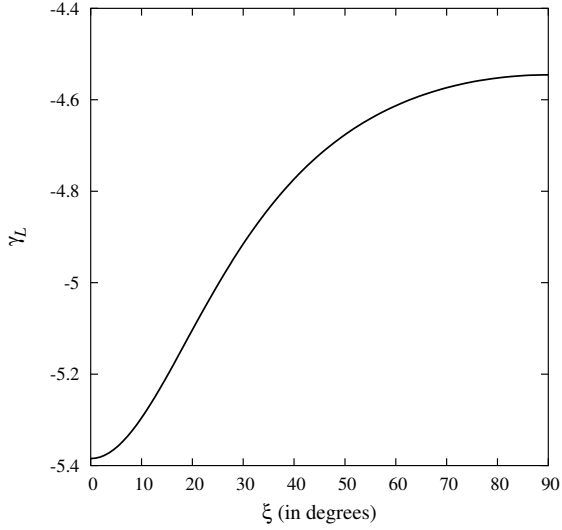} \caption{Power-law index of the
electromagnetic energy density at the light cylinder ($x=1$),
changing with the inclination angle. \label{gamm2_fig}}
\end{figure}

The value of $\gamma$
at the light cylinder, $\gamma_L\equiv \gamma(x=1)$, from equation (\ref{eq_gamm}), is
\begin{equation}
\gamma_{L}=-6+\frac{(8/9)\cos ^{2}\xi +8\sin ^{2}\xi }{(13/9)\cos
^{2}\xi +(11/2)\sin ^{2}\xi }
\label{gamm_L}
\end{equation}
and is plotted in Figure~(\ref{gamm2_fig}). The kinetic energy density scales with
$x^{-5/2}$ (see the following section). Note that for all
inclination angles $\gamma_L<-5/2$ i.e.\ the electromagnetic
energy density is steep enough to balance the kinetic energy
density of the disk at the light cylinder.

\placefigure{gamm2_fig} 

\begin{figure}[th]
\epsscale{1.0} 
\plotone{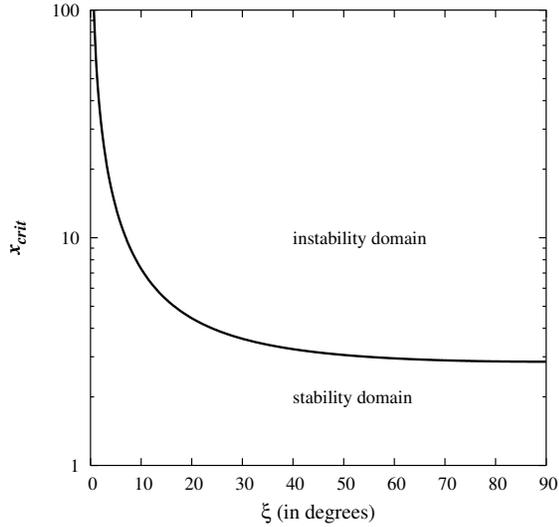} \caption{The critical radius, in
terms of light cylinder radius at which  $\gamma=-5/2$, as given
in equation~(\ref{x52}). \label{gamm52}}
\end{figure}

The radius beyond which $\gamma>-5/2$ can be found by solving
equation~(\ref{eq_gamm}) for $\gamma=-5/2$. We find this to be
\begin{equation}
x_{crit}=\frac{\sqrt{6}}{6\tan \xi }\sqrt{4+18\tan ^{2}\xi
+\sqrt{16+396\tan ^{2}\xi +954\tan ^{4}\xi }}.  \label{x52}
\end{equation}
Figure~\ref{gamm52} shows that the disk inner radius will be
stable beyond the light cylinder, for distances up to many $R_L$,
practically for all radii for aligned
($\xi=0\raisebox{1ex}{\scriptsize o}$) rotators and up to
$r_{crit}\cong 3R_L$ for orthogonal
($\xi=90\raisebox{1ex}{\scriptsize o}$) rotators.

\placefigure{gamm52}   

\section{The Inner Radius Near The Light Cylinder}
We estimate the inner radius as the electromagnetic radius $R_{em}$,
determined by
$\mathcal{E}_{em}=\mathcal{E}_{K}$ taking the conventional estimate
$\mathcal{E}_{K}=\frac{1}{2}\rho |\mathbf{v}| v_{ff}$
where $\rho |\mathbf{v}| =\dot{M}/4\pi r^{2}$ is from the
conservation of mass in spherical coordinates, $\dot{M}$ being the
mass flow rate, and $v_{ff}=\sqrt{2GM/r}$ is the free-fall
velocity. This gives
\begin{equation}
\mathcal{E}_{K} = \frac{\dot{M}\sqrt{2GM}}{8\pi
R_{L}^{5/2}}x^{-5/2}. \label{E_K}
\end{equation}
The electromagnetic radius can be found using
equations (\ref{E_tot2}) and (\ref{E_K}):
\begin{equation}
x_{em}^{7/2}=x_{A}^{7/2}\left[ \left(
1+\frac{4}{9}x_{em}^{2}\right) \cos ^{2}\xi +\left(
x_{em}^{4}+2x_{em}^{2}+\frac{5}{2}\right) \sin ^{2}\xi \right].
\label{eq_x_em}
\end{equation}
Here $x_{A}=R_{A}/R_{L}$ and
\begin{equation}
R_{A}=\left( \frac{\mu^{2}}{\dot{M}\sqrt{ 2GM}}\right) ^{2/7} \label{alfven}
\end{equation}
is the Alfv\'{e}n radius.

\begin{figure}[th]
\epsscale{0.8} 
\plotone{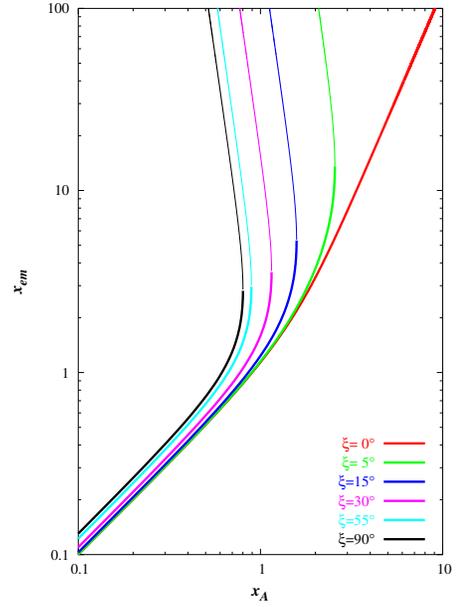} \caption{The variation of $x_{em}$
with $x_A$ as determined from Eq.~(\ref{eq_x_em}), for a variety of 
inclination angles. 
See the electronic edition of the Journal for a color version of this figure.
\label{f4}}
\end{figure}

\placefigure{f4} 

The solution of equation~(\ref{eq_x_em}) for $x_{em}(x_A)$
is given in Figure~(\ref{f4}) for a variety of  $\xi$. It is not
possible to find a solution for every $x_{A}$, corresponding to
the fact that at low mass transfer rates, the kinetic energy
density of the disk can not match the electromagnetic energy
density at any radius, and the disk will be ejected. The lowest
mass flow rate for which the disk will not be ejected corresponds
to the
maximum $x_{A}(x_{em})$ as seen in Figure~(\ref{f4}).
From $dx_{A}/dx_{em}=0$
follows the solution  (\ref{x52}) as expected. Now using this back in
equation (\ref{eq_x_em}) one finds the critical Alfv\'en
radius
\begin{equation}
x_{A,crit}=x_{crit}\left[(1+\frac{4}{9}x_{crit}^{2})\cos ^{2}\xi
+(x_{crit}^{4}+2x_{crit}^{2}+\frac{5}{2})\sin ^{2}\xi
\right]^{-2/7} \label{xac}
\end{equation}
corresponding to the single solution $x_{em}=x_{crit}$ and to the critical mass flow rate
\begin{equation}
\dot{M}_{crit} =\frac{\mu ^{2}}{\sqrt{2GM}R_{L}^{7/2}}x_{A,crit}^{-7/2}. \label{mdotcrit}
\end{equation}
For higher mass flow rates there will be two solutions for $x_{em}$
the smaller of which is the stable one. This stable solution is
beyond the light cylinder for a range of mass flow rates
$\dot{M}_{crit}<\dot{M}<\dot{M}_{L}$ where $\dot{M}_{L}$ is the
flow rate for which the inner radius will be at the light
cylinder, $x_{em}=1$. This will happen when $x_{A}$ attains the
value
\begin{equation}
x_{A,L}=\left( \frac{13}{9}\cos ^{2}\xi +\frac{11}{2}\sin ^{2}\xi
\right)^{-2/7} \label{xal}
\end{equation}
as can be obtained from equation~(\ref{eq_x_em}).
Hence
\begin{equation}
\dot{M}_{L} =\frac{\mu ^{2}}{\sqrt{2GM}R_{L}^{7/2}}x_{A,L}^{-7/2}  \label{Mdot_L}
\end{equation}
is obtained.
Note that both $\dot{M}_{crit}$ and $\dot{M}_{L}$ are strongly
dependent on the period and the magnetic moment of the neutron
star ($\dot{M}_{L} \propto \dot{M}_{crit} \propto P^{-7/2}\mu^2$)
and their ratio is only dependent on the inclination angle $\xi$.
The range of mass flow rates and inner disk radii for which the disk
is beyond the light cylinder, and stable, is smallest for
$\xi=90\raisebox{1ex}{\scriptsize o}$; for which $R_{crit} \cong 3 R_L$ and
$\dot{M}_{crit} = 0.4 \dot{M}_L$.

Within the model the disk will
survive beyond the light cylinder, even if the radio pulsar
activity turns on. As long as $\dot{M}$ remains larger than
$\dot{M}_{crit}$ (i.e. $R_A<R_{A,crit}$), the disk will not be
ejected, and the radio pulsar may turn off again when $\dot{M}$
increases to $\dot{M}_L$, switching back to the propeller phase. 
The threshold for quenching the radio pulsar
will reflect loading of all closed field lines that lead to the inner gap
with plasma from
the disk. This does not necessarily happen at $R_{em}=R_L$. If the
transition to radio pulsar activity happens at some radius
$R_{t}(\xi)<R_L$, then the transient radio pulsar phase with the
disk not ejected prevails for $\dot{M}<\dot{M}_{t}$,
corresponding to $R_{em}>R_{t}$.

The scaling we have used which
reduces to the Alfv\'{e}n radius for $r \ll R_L$ presumes a spherical accretion flow.
The \citet{SS73} solutions for thin Keplerian disks
asymptotically imply that the density of the gas in the disk scales  as
$\rho \propto r^{-15/8}$.
Hence $\mathcal{E}_K\propto \rho v_{K}^{2} \propto \rho
r^{-1}\propto r^{-23/8}$. This scaling ($\gamma=-23/8=-2.875$) is slightly
more steep than what we have used $\gamma=-5/2$. For a Shakura-Sunyaev thin disk,
our estimate for the critical inner radius will be somewhat smaller than the estimate given above.
From Eq.~(\ref{x52}) the minimum critical radius ($\xi=90\raisebox{1ex}{\scriptsize o}$)
is $x_{crit,min}=2.85$ while for $\gamma=-2.875$ we would obtain $x_{crit,min}=2.13$.

\section{Applications}

\subsection{SAX J1808.4-3658, Other Millisecond X-ray Pulsars and Transients}

\begin{figure}[h]
\epsscale{1.1} 
\plotone{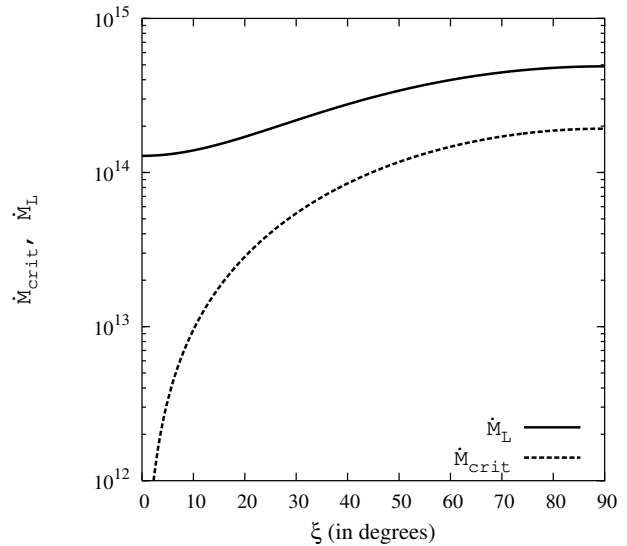} \caption{The mass flow rate range
for SAX J1808.4--3658 for which the inner radius is outside the
light cylinder while the disk is not swept away by the radiation
pressure. The \emph{solid line} corresponds to $\dot{M}_L$ (eqn.~\ref{Mdot_L})
and the \emph{dashed line} corresponds to $\dot{M}_{crit}$ (eqn.~\ref{mdotcrit}). 
The domain between the curves corresponds to the allowed mass flow range 
for which the inner radius is outside the light cylinder while the disk is not disrupted.
The observed luminosity may be much lower than that would
correspond to these mass flow rates because the
centrifugal barrier will not allow all inflowing mass to accrete
onto the neutron star. For small inclination angles the allowed
mass flow rate range increases. \label{sax1808}}
\end{figure}

The rotation periods of accretion driven millisecond pulsars \citep{wijnands04} 
range between 2.3 ms to 5.4 ms.
We take the first discovered \citep{wijnands98} accretion driven millisecond pulsar
SAX J1808.4-3658 with a rotation period $P = 2.5$ ms 
as an example of this class. The
upper limit on the magnetic field of the neutron star is $5 \times
10^8$ Gauss \citep{disalvo}. We assume the magnetic moment to
be $\mu= 3 \times10^{26}$ Gauss cm$^3$. Using these parameters in equations
(\ref{mdotcrit}) and (\ref{Mdot_L}) we obtain the results shown
in Figure~\ref{sax1808} for \objectname{SAX J1808.4-3658}. For mass flow rates greater than  $10^{13}$ g s$^{-1}$ for 
an inclination angle $ \xi=10 \raisebox{1ex}{\scriptsize o}$ and greater than $2 \times 10^{14}$ g s$^{-1}$ for 
an inclination angle $ \xi=90 \raisebox{1ex}{\scriptsize o}$, the source can sustain a disk beyond the light cylinder 
and thus can be in a rotation powered pulsar phase. The region between the two curves
represents the mass inflow rate for which the electromagnetic radius
is larger than the light cylinder while the disk can survive the pulsar activity. At mass inflow rates $\dot{M}_L$ 
greater than a few 10$^{14}$ g s$^{-1}$ for all inclination angles, the disk inner radius will protrude into the light cylinder, 
and magnetosperic pulsar activity will be quenched by the presence of the disk to the extent that all field lines are 
loaded with plasma from the disk. The source will probably be in a propeller phase with only a 
fraction of the mass inflow possibly being accreted while the rest is diverted into an outflow. 
When the inner radius of the disk is of the order of the corotation radius, accretion of the full mass flow rate is expected. 
We estimate mass inflow rates of the order of 10$^{16}$ g s$^{-1}$ for this transition for all inclination angles.  
The maximum luminosity of \objectname{SAX J1808.4-3658} in outburst is 
$\sim 2\times 10^{35}$ erg s$^{-1}$, which corresponds to a mass accretion rate of 10$^{14}$ g s$^{-1}$. 
The discrepancy is probably related to our scaling of the disk inner radius with  
the Alfv\'en radius, and the uncertainties in the determination of the Alfv\'en radius as well as to the simplifications 
of the present model in terms of the vacuum dipole solutions. 
In quiescence the luminosity drops to $\sim 10^{32}$ erg s$^{-1}$ \citep{campana02}. 
It is possible that the mass inflow rate in the disk is large enough that
the inner radius of the disk is inside the light cylinder but only a very small fraction
($\lesssim 0.1$ percent) of this inflowing mass accretes onto the neutron
star, producing the observed X-ray luminosity. \citet{burderi03} argued that the irradiation of the companion by the switched
on magneto-dipole rotator, in the quiescent stage, can explain the 
modulation of the flux in the optical (see also \citet{campana04}). If the source is in the propeller stage accretion of only a small fraction of the disk inflow 
through a limited bunch of field lines will probably allow the magnetospheric gaps and pulsar radiation to survive. 
This would be easier for the outer gaps. Magnetospheric voltages in millisecond pulsars are of the order of those in the Vela pulsar, 
so optical, X and gamma ray pulsar activity may be possible if  the outer gap can survive. For small inclination angles accretion 
near the magnetic polar caps would 
also be near the rotational pole and therefore avoiding the centrifugal barrier. For paths avoiding the inner gap 
even radio pulsar activity might survive in the propeller phase, with the disk protruding inside the light cylinder. 
Depending on the beaming geometry searches for radio and high energy pulsar activity might yield very interesting results. 
\objectname{SAX J1808.4-3658} was selected for this discussion as it is the millisecond X-ray pulsar with the most detailed information 
\citep{wijnands04}. This discussion applies to the other millisecond X-ray pulsars also.

\placefigure{sax1808}

Other  transient low mass  X-ray binaries (LMXBs) may also be in phases that include excursions 
of the inner disk towards and beyond the light cylinder and back. As an example, the luminosity 
of Soft X-ray Transient \objectname{Aquila X-1} 
was observed \citet{campana} to decay from $\sim 10^{36}$ erg s$^{-1}$ to $\sim 10^{33}$ erg s$^{-1}$, 
from outburst to the 
quiescent stage. As the system in the quiescent stage is likely to be in the propeller mode, 
the observed luminosity of $\sim 10^{33}$ erg s$^{-1}$ reflects a small fraction of the inflowing 
mass that accretes.  
The present discussion adds to the possibilities that of occasional rotation 
powered pulsar activity. 
The uncertainty in rotation period and magnetic field does not allow a more detailed 
discussion in terms of our model. 
Further, the majority of LMXBs do not exhibit rotation periods in their accretion 
and propeller phases for reasons of selection 
in the source, and this is probably also the case for most transients.

\subsection{Implications for Fallback Disks}

Neutron stars are born in supernova explosions. It is a
possibility that some of the ejected matter fails to achieve the
escape velocity and falls back \citep{woos02}. As the progenitor
is rotating,  the fallback matter might have enough angular
momentum to settle into a disk  \citep{mineshige97}. The idea of
fallback disks around pulsars dates back to \citet{MD81}. The
fallback disk model \citep{CHN,alpar01,marsden01a,EA03}
of anomalous X-ray pulsars \citep{mereghetti,kaspi}
renewed interest in possible implications for radio pulsars
\citep{MD81,marsden01b,marsden02,menou01,AAY01}. More recently
\citet{BP04} suggested that the observed jets of Crab and Vela
may be collimated by fallback disks.

Young neutron stars might follow a variety of evolutionary paths
depending on the initial mass of their fallback disks
\citep{alpar01}. The evolutionary stages are determined
by the mode of interaction discussed in \S 1. As the mass of a fallback
disk is not replenished, mass flow rate in the disk declines and the
inner radius of the disk, as a consequence, moves out. As the neutron star is
initially spinning down with propeller torques, the light cylinder is also
moving out, though less rapidly than the inner radius. At some stage
the inner radius will catch up with the light cylinder.

\citet{menou01} and \citet{AAY01} assumed that the
inner radius of  fallback disks
tracks the light cylinder radius. In such a model
the disk can assist  the magnetic dipole radiation torque
without quenching the radio pulsar mechanism.
These models would not work \citep{li02} if the disruption of the disk
takes place \citep{shv70b} as soon as its inner edge reaches the light
cylinder from within the
near zone. The present work shows that such a disk need not be necessarily
disrupted till its inner radius moves to a few times the light cylinder radius.
Whether the inner radius can actually track the light cylinder for a
long epoch, rather than moving beyond the light cylinder into the radiation zone,
will be studied in a subsequent paper.

\section{Conclusion}

We have shown that the electromagnetic energy density of a
rotating dipole makes a rather broad transition across the light
cylinder from the near zone dipole magnetic field to the radiation
zone, the transition being broader for small inclination angles, $\xi$.
For power-law indices $\gamma$ for the $r$ dependence of the
electromagnetic energy density, $\gamma>-5/2$, the regime in which
the radiation pressure would eject the disk, is not attained until
 $r$ exceeds a few $R_L$ depending
on the inclination angle between the spin and magnetic axes (see
eqn.~(\ref{x52}) and Figure~\ref{gamm52}). Disks that start their
evolution with the inner radius far inside the light cylinder may
evolve to the situation where the inner disk radius reaches the
light cylinder. In this case and even further the electromagnetic
stresses of the fields generated by the neutron star can balance the
material stresses from the disk. Near the light cylinder the
power-law index of the electromagnetic energy density $\gamma
_{L}$ varies between $-5.4$ (for $\xi =0 \raisebox{1ex}{\scriptsize o}$) and $-4.5$ (for $\xi
=90 \raisebox{1ex}{\scriptsize o}$) (see Eq.~(\ref{gamm_L}) and Fig~\ref{gamm2_fig}).
Assuming a thin Shakura-Sunyaev disk with $\mathcal{E}_K \propto
r^{-23/8}$ does not change these conclusions qualitatively and a
stable inner radius can be found at least up to $\cong 2$ times
the light cylinder radius for orthogonal rotators while this
critical radius inside which stable equilibrium can be found may
extend up to a few ten times the light cylinder for small
inclination angles.

When the inner radius of the disk is outside the light cylinder, the pulsar
activity is likely to turn on.The presence of the disk may effect the coherent radio emission.
The magnetic dipole radiation torque will act on the neutron star even if a radio pulsar is not observed. 
Sources on their evolutionary path to higher or lower mass inflow rates will make a transition from or to a 
stage with a disk inner radius stablely placed in the radiation zone, and with possible rotation powered activity. 
Millisecond X-ray pulsars are likely examples of late stages in the evolution of LMXBs into millisecond radio pulsars 
through spin-up by accretion \citep{acrs82,rs82}.
Such sources might hover around the transition, exhibiting transient behaviour. The stable presence of the disk outside the 
light cylinder for a wide range of mass inflow rates makes repeated transitions and sustained transient behaviour possible 
not only around the transition between propeller and accretion phases, but also around the transition of the inner 
radius of the disk across the light cylinder. 

\acknowledgements
We thank \"Unal Ertan, Rashid Sunyaev and Ed van den Heuvel for discussions. KYE thanks Nihal Ercan for her encouragement. 
This work was supported by Sabanc\i\ University
Astrophysics and Space Forum, by the High Energy Astrophysics Working 
Group of T\"{U}B\.{I}TAK
(The Scientific and Technical Research Council of Turkey)
and by the Turkish Academy of Sciences for MAA.

\appendix

\section{The Deutsch (1955) Solutions}

For an inclination angle $\xi $, equatorial magnetic field
strength $B_{0}$,
and stellar radius $R_{\ast }$, the solutions in spherical coordinates ($%
r,\theta ,\phi $) are the following (\citet{ML99}, Eqns. 89 \&
90):
\begin{eqnarray}
B_{r} &=&2B_{0}\frac{R_{\ast }^{3}}{r^{3}}\left\{ \cos \xi \cos
\theta +\sin \xi \sin \theta \left[ d_{1}\cos \psi +d_{2}\sin \psi
\right] \right\} ,
\nonumber \\
B_{\theta } &=&B_{0}\frac{R_{\ast }^{3}}{r^{3}}\left\{
\cos \xi \sin \theta
-\sin \xi \cos \theta \left[ (q_{1}+d_{3})\cos \psi +(q_{2}+d_{4})\sin \psi %
\right]
\right\} ,  \label{gen1} \\
B_{\phi } &=&B_{0}\frac{R_{\ast }^{3}}{r^{3}}\sin \xi \left\{
-\left[
q_{2}\cos 2\theta +d_{4}\right] \cos \psi +\left[ q_{1}\cos 2\theta +d_{3}%
\right] \sin \psi \right\}   \nonumber
\end{eqnarray}%
and%
\begin{eqnarray}
E_{r} &=&\frac{E_{0}}{c}\frac{\alpha ^{2}}{x^{2}}\left\{
\frac{2}{3}\cos \xi +\frac{\alpha ^{2}}{x^{2}}\cos \xi (1-3\cos
^{2}\theta )
-\frac{3}{x^{2}}\sin \xi \sin 2\theta \lbrack q_{1}\cos \psi
+q_{2}\sin \psi
]
\right\} ,  \nonumber \\
E_{\theta } &=&\frac{E_{0}}{c}\frac{\alpha ^{2}}{x^{2}}\left\{
-\frac{\alpha ^{2}}{x^{2}}\cos \xi \sin 2\theta +\sin \xi \lbrack
(q_{3}\cos
2\theta -d_{1})\cos \psi
+(q_{4}\cos 2\theta -d_{2})\sin \psi ]%
\right\} ,  \label{gen2} \\
E_{\phi } &=&\frac{E_{0}}{c}\frac{\alpha ^{2}}{x^{2}}\sin \xi \cos
\theta \left[ (q_{4}-d_{2})\cos \psi -(q_{3}-d_{1})\sin \psi
\right]   \nonumber
\end{eqnarray}
Here
\begin{equation}
\psi =\phi -\Omega t+x-\alpha   \label{psi1}
\end{equation}
where $\Omega $ is the angular velocity of the star,
\begin{equation}
x =\frac{r}{R_{L}}=\frac{r\Omega }{c}  \label{x}
\end{equation}
and
\begin{equation}
\alpha =\frac{R_{\ast }}{R_{L}}=\frac{R_{\ast }\Omega
}{c}=\allowbreak 6.\,\allowbreak 28\times 10^{-3}\left(
\frac{P}{100ms}\right) ^{-1}. \label{alfa}
\end{equation}
For usual rotation rates $\alpha \ll 1$. Moreover,
\begin{equation}
E_{0} =\Omega R_{\ast }B_{0}=c\alpha B_{0}  \label{E_0}
\end{equation}
and the non-constant coefficients are
\begin{eqnarray}
d_{1} &=&\frac{\alpha x+1}{\alpha ^{2}+1},\qquad d_{2}=\frac{x-\alpha }{
\alpha ^{2}+1},  \nonumber \\
d_{3} &=&\frac{1+\alpha x-x^{2}}{\alpha ^{2}+1},\qquad d_{4}=\frac{
(x^{2}-1)\alpha +x}{\alpha ^{2}+1}  \label{ds}
\end{eqnarray}
and

\begin{eqnarray}
q_{1} &=&\frac{3x(6\alpha ^{3}-\alpha ^{5})+(3-x^{2})(6\alpha
^{2}-3\alpha
^{4})}{\alpha ^{6}-3\alpha ^{4}+36},  \nonumber \\
q_{2} &=&\frac{(3-x^{2})(\alpha ^{5}-6\alpha ^{3})+3x(6\alpha
^{2}-3\alpha
^{4})}{\alpha ^{6}-3\alpha ^{4}+36}, \\
q_{3} &=&\frac{(x^{3}-6x)(\alpha ^{5}-6\alpha
^{3})+(6-3x^{2})(6\alpha
^{2}-3\alpha ^{4})}{x^{2}(\alpha ^{6}-3\alpha ^{4}+36)},  \label{qs} \\
q_{4} &=&\frac{(6-3x^{2})(\alpha ^{5}-6\alpha
^{3})+(6x-x^{3})(6\alpha ^{2}-3\alpha ^{4})}{x^{2}(\alpha
^{6}-3\alpha ^{4}+36)}.
\end{eqnarray}

We assume the disk to be at the $\theta =\pi /2$ plane
where the field components given in
(\ref{gen1}) and (\ref{gen2}) simplify to
\begin{eqnarray}
B_{r} &=&\frac{2\mu }{R_{L}^{3}x^{3}}\sin \xi \left( d_{1}\cos
\psi
+d_{2}\sin \psi \right) ,  \nonumber \\
B_{\theta } &=&\frac{\mu }{R_{L}^{3}x^{3}}\cos \xi ,  \label{pl1} \\
B_{\phi } &=&\frac{\mu }{R_{L}^{3}x^{3}}\sin \xi \left[ \left(
q_{2}-d_{4}\right) \cos \psi +\left( d_{3}-q_{1}\right) \sin \psi
\right] \nonumber
\end{eqnarray}
where $\mu =B_{0}R_{\ast }^{3}$ and using (\ref{E_0}) in (\ref
{gen2})
\begin{eqnarray}
E_{r} &=&\frac{\mu }{R_{L}^{3}x^{2}}\cos \xi \left(
\frac{2}{3}+\frac{\alpha
^{2}}{x^{2}}\right) ,  \nonumber \\
E_{\theta } &=&-\frac{\mu }{R_{L}^{3}x^{2}}\sin \xi \lbrack
(q_{3}+d_{1})\cos \psi +(q_{4}+d_{2})\sin \psi ],  \label{pl2} \\
E_{\phi } &=&0.  \nonumber
\end{eqnarray}
We obtain the radial component of the Poynting vector
$\mathbf{S}=(c/4\pi ) \mathbf{E\times B}$ at the disk plane
($\theta=\pi/2$) from equations (\ref{pl1}) and (\ref{pl2}).
Averaging $S_r=\mathbf{S \cdot \hat{r}}$ over a period, we obtain
\begin{equation}
\langle \frac{\mathbf{S}_{r}}{c}\rangle =\frac{\mu ^{2}}{8\pi R_{L}^{6}}%
\frac{\sin ^{2}\xi }{x^{5}}s(x) \label{S_r_App}
\end{equation}
where
\begin{equation}
s(x)=-[(q_{3}+d_{1})\left( q_{2}-d_{4}\right) +(q_{4}+d_{2})\left(
d_{3}-q_{1}\right) ]
\end{equation}
This can be approximated as
\begin{equation}
s(x)\simeq
\begin{cases}
\frac{3x^{5}+2\alpha ^{2}\allowbreak x^{3}+\alpha ^{5}}{3x^{2}} & \text{for $
x$}\sim \alpha \text{;} \\
x^{3} & \text{for $x\gtrsim $ a few }\alpha \text{.}
\end{cases}
\label{As}
\end{equation}

We average the square of the fields over one stellar period $P$ as $\langle
G_{i}^{2}\rangle =(1/P)\int_{0}^{P}G_{i}^{2}dt$ where $G_{i}$ is any
component of the electric or magnetic field.
\begin{eqnarray}
\mathcal{E}_{m} &=&\frac{\langle B^{2}\rangle }{8\pi } \\
&=&\frac{\mu ^{2}}{8\pi R_{L}^{6}}x^{-6}\left( \cos ^{2}\xi +\frac{1}{2}%
f_{1}(x)\sin ^{2}\xi \right)
\end{eqnarray}
where
\begin{equation}
f_{1}(x)=\left( q_{2}-d_{4}\right) ^{2}+\left( d_{3}-q_{1}\right)
^{2}+4d_{1}^{2}+4d_{2}^{2}
\end{equation}
In the limit $\alpha \ll 1$, this  simplifies as $f_{1}(x)=x^{4}+3x^{2}+5$
and we obtain
\begin{equation}
\mathcal{E}_{m}\simeq \frac{\mu ^{2}}{8\pi R_{L}^{6}}x^{-6}\left[ \cos
^{2}\xi +\frac{1}{2}\left( x^{4}+3x^{2}+5\right) \sin ^{2}\xi \right]
\end{equation}
Similarly the electric energy density is
\begin{eqnarray}
\mathcal{E}_{e} &=&\frac{\langle E^{2}\rangle }{8\pi } \\
&=&\frac{\mu ^{2}}{8\pi R_{L}^{6}}x^{-4}\left[ \cos ^{2}\xi \left( \frac{2}{3}
+\frac{\alpha ^{2}}{x^{2}}\right) ^{2}+\frac{1}{2}f_{2}(x)\sin ^{2}\xi
\right]
\end{eqnarray}
where
\begin{equation}
f_{2}(x)=(q_{3}+d_{1})^{2}+(q_{4}+d_{2})^{2}
\end{equation}
and in the limit $\alpha \ll 1$ this simplifies as $f_{2}(x)=x^{2}+1$ and we
obtain
\begin{equation}
\mathcal{E}_{e}\simeq \frac{\mu ^{2}}{8\pi R_{L}^{6}}x^{-4}\left[ \frac{4}{9}
\cos ^{2}\xi +\frac{1}{2}\left( x^{2}+1\right) \sin ^{2}\xi \right]
\end{equation}

\clearpage

\end{document}